\documentclass[usenatbib,onecolumn]{mn2e}
\usepackage{newtxtext,newtxmath}
\usepackage{lipsum}
\usepackage[T1]{fontenc}
\usepackage{ae,aecompl}
\usepackage{hyperref}
\usepackage{graphicx}	
\usepackage{amsmath}	
\usepackage{amssymb}	
\usepackage{bm}	
\usepackage{caption}
\usepackage{multirow}
\usepackage{multicol}
\usepackage{cuted}
\usepackage{mathtools}
\usepackage{longtable}

\title[Epsilon Indi]{Detection of the closest Jovian exoplanet in the Epsilon Indi triple system}
\author[F. Feng et al.]
{F. Feng$^{1}$\thanks{E-mail: f.feng@herts.ac.uk or fengfabo@gmail.com}, M. Tuomi$^{1}$, H. R. A.
  Jones$^{1}$\\
$^{1}$Centre for Astrophysics Research, University of Hertfordshire, College
Lane, AL10 9AB, Hatfield, UK}
\date{\today}

\begin{document}
\maketitle
\begin{abstract}
  We confirm the trend in the radial velocity data for Epsilon Indi A suggesting a long-period planetary companion and find significant curvature is present, sufficient to quantify Epsilon Indi Ab as a cold Jupiter with a minimum mass of $2.71_{-0.44}^{+2.19}~M_{\rm Jup}$ on a nearly circular orbit with a semi-major axis of $12.82_{-0.71}^{+4.18}$\,au and an orbital period of $52.62_{-4.12}^{+27.70}$\,yr. We also identify other significant signals in the radial velocity data. We investigate a variety of spectral diagnostics and interpret these signals as arising from activity-induced radial velocity variations. In particular, the 2500 and 278\,d signals are caused by magnetic cycles. While a planetary signal might be present in the 17.8 d signal, the origin of 17.8 and 11 d signals are most easily interpreted as arising in the rotation of the star with a period of about 35\,d. We find that traditional activity indicators have a variety of sensitivities. In particular, the sodium lines and CaHK index are sensitive to all activity-induced signals. The line bisector measurement is sensitive to stellar rotation signal while H$\alpha$ is sensitive to the secondary magnetic cycle. In general, because of their different sensitivities these activity indicators introduce extra noise if included in the noise model whereas differential RVs provide a robust proxy to remove wavelength-dependent noise efficiently. Based on these analyses, we propose an activity diagnostics procedure for the detection of low amplitude signals in high precision radial velocity data. Thus the Epsilon Indi system comprises of at least Epsilon Indi A, Ab as well as a long period brown dwarf binary Ba and Bb; so it provides a benchmark case for our understanding of the formation of gas giants and brown dwarfs.
\end{abstract}
\section{Introduction}     \label{sec:introduction}
One of the goals of exoplanet searches is to discover Earth-like planets around Sun-like stars. So called Earth analogs located in the temperate zone of a Sun-like star (where liquid water can exist) induce a stellar radial velocity (RV) variation of about 0.1\,m/s, which is below the detection limit of current spectrographs. Notably, Earth-size planets in the temperate zone of an M-dwarf can cause an RV variation of a few m/s and are relatively easier to detect (e.g. Proxima Centauri b; \citealt{anglada16}). However, such planets are close to the star, thus are tidally locked to the star and their atmospheres suffer from strong erosion by high incident ultraviolet stellar fluxes \citep{lingam17}. In between G and M dwarfs, K dwarfs may represent excellent targets for the detection of Earth analogs through the RV method. 

Epsilon Indi A is a good candidate for searching Earth analogs. Epsilon Indi A (HIP 108870, HR 8387, HD 209100, GJ 845) is a young (1.4 Gyr, \citealt{bonfanti14}) nearby K2V star (3.62\,pc according to \citealt{leeuwen07}) with a mass of 0.762$\pm$0.038\,$M_\odot$ \citep{demory09}, and a luminosity of 0.22\,$L_\odot$. This star is also accompanied by a binary brown dwarf with a separation of about 1459\,au \citep{scholz03}. The trend observed in the RV data is much larger than from the brown dwarfs and supports the existence of another companion with a period longer than 30\,years \citep{endl02,zechmeister13}. 

To find small Keplerian signals and to investigate the previously proposed long period companion around Epsilon Indi A, we analyze the data from \cite{zechmeister13} and the recent HARPS data in the ESO archive in the Bayesian framework. With new data and noise modelling techniques there is the potential for detection of weak signals (e.g. \citealt{feng16, feng17b}). Using the {\small Agatha} software \citep{feng17a}, we compare noise models to find the so-called Goldilocks model \citep{feng16}. We also calculate the Bayes factor periodograms (BFPs) to test the sensitivity of signals to the choice of noise models. These signals are further diagnosed by calculating the BFPs for various noise proxies and their residuals.

This paper is structured as follows. First, we introduce the data in section \ref{sec:data}. Then we describe the statistical and numerical methods for the analysis of RV data and constrain the long period signal present in the data. In section \ref{sec:signal}., we concentrate on the HARPS data and constrain the activity signals using posterior samplings. In section \ref{sec:diagnose} we present the general efficacy of differential RVs on a larger dataset of stars and formulate a general diagnostic procedure for distinguishing between activity and planetary signals. Finally, we discuss and conclude in section \ref{sec:conclusion}. 

\section{Data}\label{sec:data}
\begin{figure*}
  \begin{center}
\includegraphics[scale=0.55]{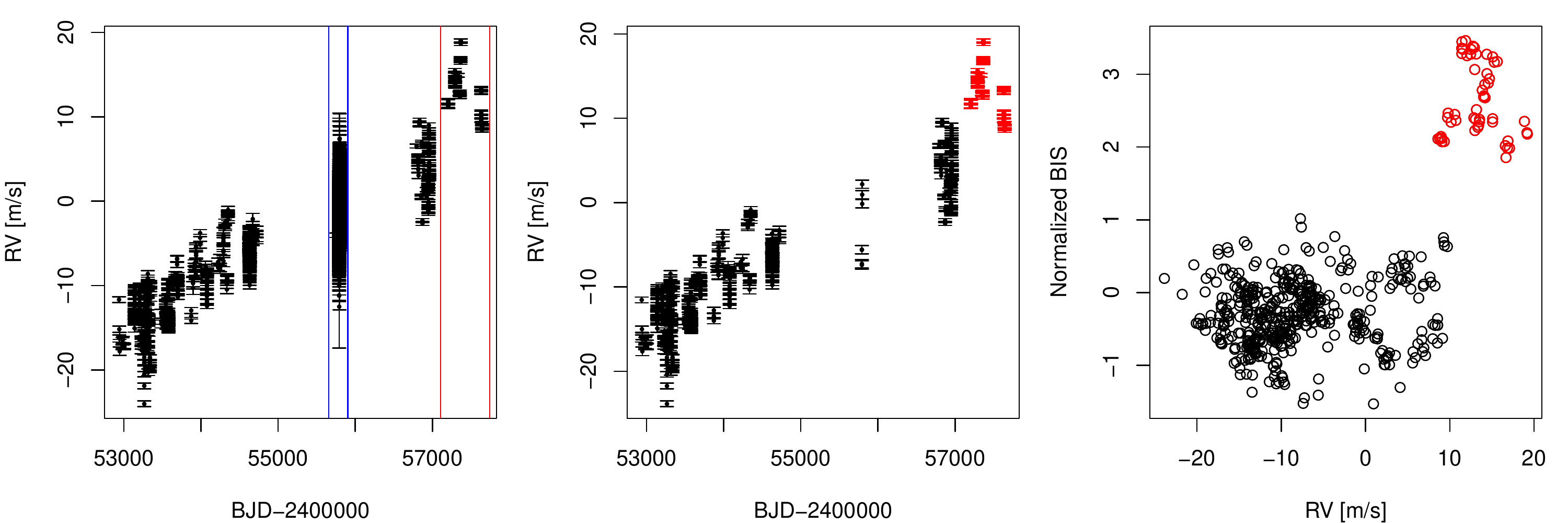}
  \caption{HARPS data of Epsilon Indi A. Left panel: the blue vertical lines enclose the high cadence data while the red lines enclose the data with high BIS. All these data points define the HH data set. Middle panel: the HL RVs including all HH RVs but excluding the high cadence RVs with low signal to noise ratio. The RVs with high BIS are shown in red. These abnormal RVs (or postCF RVs) plotted in red are removed from the HL set to define the HC data set. Right panel: correlation between BISs and RVs of the HL set. Similar to the middle panel, the postCF points are shown in red. }
\label{fig:data}
\end{center}
\end{figure*}

We obtain the HARPS data by processing the spectra in the ESO archive using the TERRA algorithm \citep{anglada12}. The HARPS data also include various noise proxies courtesy of the HARPS pipeline, including R$'_{\rm HK}$ (or CaHK index), Bisector span (BIS), full width half maximum of spectra lines (FWHM). These are supplemented by TERRA-generated indices including intensity of the H-alpha line (H-alpha), indices from sodium lines (NaD1 and NaD2), and the differential RVs which are derived from the 72 echelle orders of HARPS spectrum \citep{feng17b}. For example, the 3AP2-1 differential RV set is the difference between the 3AP2 and 3AP1 aperture data sets. We divide the 72 echelle orders evenly into three divisions, and the 3AP1 and 3AP2 aperture data sets are the averaged RVs of the first and second divisions. In the left-hand panel of Fig. \ref{fig:data}, we show all the HARPS RVs including high cadence epochs (called ``HH''; 4198 points). We note that the RVs from JD2455790 to JD2455805 are measured with high cadence, leading to 3636 RVs spread over two weeks. Such high-cadence data were obtained to study high frequency stellar oscillations and most of them are measured with a considerably lower signal to noise ratio than the rest of the data. To remove the stellar and instrumental systematics, we define the HL data set by excluding RVs with signal to noise ratio less than 110 and plot them in the middle-panel of Fig. \ref{fig:data}. This dataset excluding most of the high cadence data with low signal to noise is called ``HL'' and consists of 518 points. We further remove the RVs with high BIS from the HL set with the further caveat that  we are aware of the ill-determined offset for post-2015 data \citep{curto15} to define a more conservative data set called ``HC'' of 465 points by using the correlation between HL RVs and BIS based on the right-hand panel of Fig. \ref{fig:data}.  These excluded RVs are due to a change of fibre in HARPS and are named ``postCF''. An offset parameter is needed to combine these data points with HC. To avoid potential degeneracy between this offset and Keplerian signals , we use HC to investigate short-period signals and use both HC and HL to constrain the trend. Moreover, we also use the CES long camera (LC) and very long camera (VLC) data sets from \cite{zechmeister13} to constrain the trend. 

\section{Detection of a cold Jupiter}\label{sec:detection}
\begin{figure*}
 \begin{center}
\includegraphics[scale=0.43]{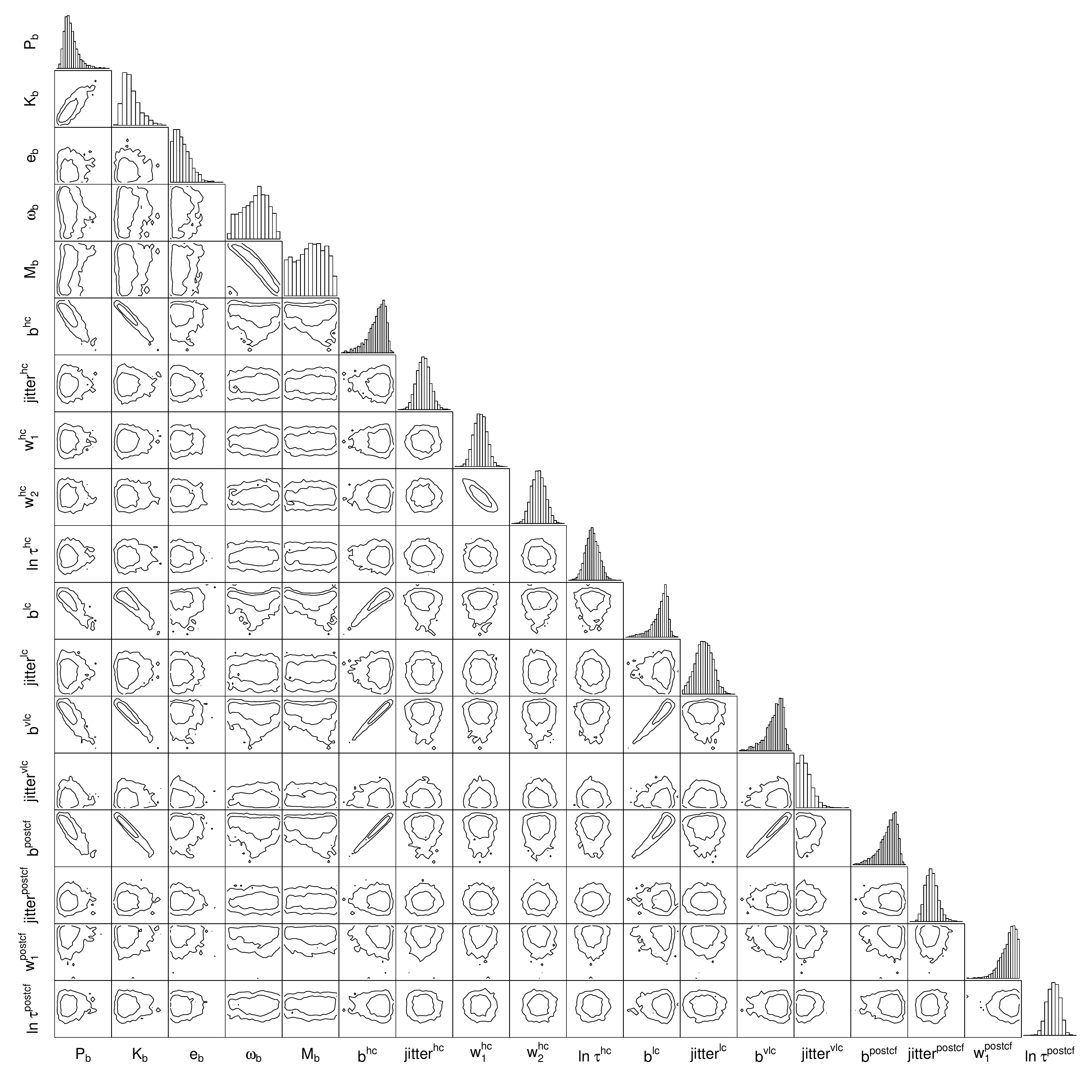}
 \caption{Two-dimensional posterior distributions for all model parameters. The first five parameters are for the Keplerian signal, $b$ is offset, $w$ and $\ln{\tau}$ are respectively the amplitude and time scale of the MA model. The contour levels correspond to the significance of 68\% and 95\%. }
\label{fig:post}
\end{center}
\end{figure*}

A significant RV trend has been found in the CES and HARPS data sets \citep{endl02,zechmeister13}, suggesting a long-period planetary companion to Epsilon Indi A. To constrain this RV trend better by including newer HARPS data, we use the CES LC and VLC data sets in combination with the HC and postCF sets. The LC and VLC data sets are corrected by accounting for the 1.84\,m\,$s^{-1}$/\,yr acceleration caused by the proper motion.  We model the trend in the combined data using one Keplerian function, and model the noise in LC, VLC, HC, and postCF using the white noise, white noise, MA(2), and MA(1) models, respectively, based on Bayesian model comparison \citep{feng17a}. We also vary the offsets between data sets and adopt the prior distributions for all parameters from \cite{feng16}. Specifically, we use a Gaussian prior distribution for eccentricity, a log uniform distribution for time scale parameters, and a uniform distribution for other parameters. Based on adapted MCMC samplings \citep{haario06}, we find a significant signal at a period of $19205.51_{-1502.23}^{+10111.37}$\,d or $52.62_{-4.12}^{+27.70}$\,yr,
which is well constrained by MCMC samplings, see Figure \ref{fig:post}.

\begin{figure}
 \begin{center}
\includegraphics[scale=0.6]{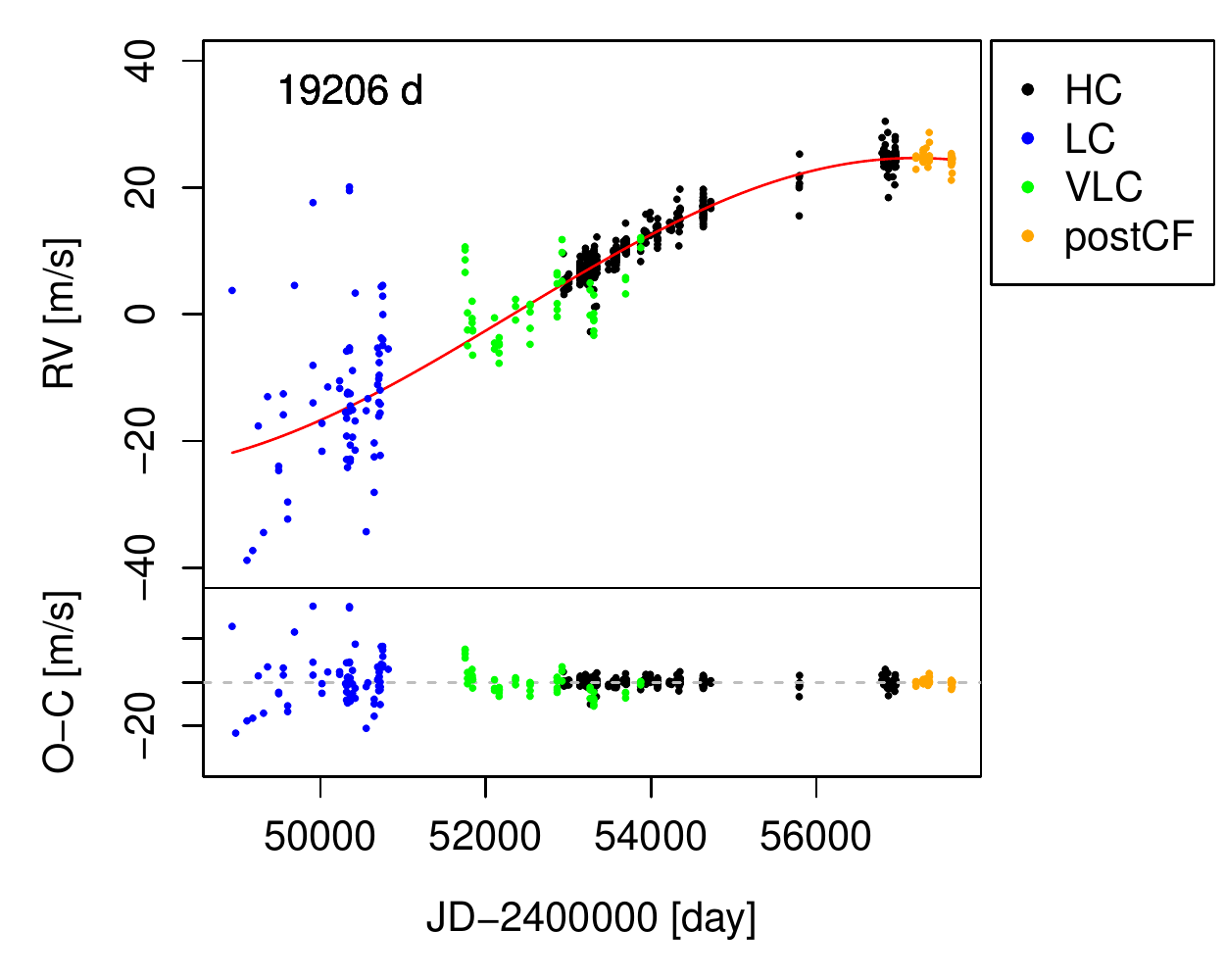}
 \caption{Best-fit for the data sets of CES-LC, CES-VLC, HC, and postCF. The offset and correlated noise are subtracted from the data. The red line denotes the Keplerian signal with a period of 52.62\,yr.  }
\label{fig:fit}
\end{center}
\end{figure}

Since this trend is not found in noise proxies, it is likely to be caused by a wide-orbit planet with a minimum mass $m\sin{i}=2.71_{-0.44}^{+2.19}~M_{\rm Jup}$, semi-major axis $a=12.82_{-0.71}^{+4.18}$, semi-amplitude $K=24.67_{-3.50}^{+14.28}$\,m/s, eccentricity of $e=0.01_{-0.01}^{+0.12}$, argument of periapsis $\omega_0=88.56_{-185.33}^{+23.28}$\,deg, and mean anomaly of $M_0=117.80_{-49.41}^{+182.14}$\,deg. This solution for Epsilon Indi Ab is shown in Fig. \ref{fig:fit}. The offset between VLC and LC sets is 5.31\,m/s, which is within the uncertainty of 8\,m/s determined from a sample of VLC and LC sets \citep{zechmeister13}. The offset between the postCF and HC set is 9.65\,m/s, consistent with the offset between pre-and-post fibre exchange for a K2 star given by \cite{curto15}. However, the mass and semi-major axis determined in this work is not consistent with the ones given by \cite{janson09}, who adopt a 2.6\,m\,$s^{-1}$/\,yr acceleration determined from epochs earlier than JD2455000 (or June, 2009). With the benefit of the number of years of precise HARPS epochs we are in a position to determine a much more modest slope value of 0.4\,m\,s$^{-1}{\rm yr}^{-1}$ for recent epochs. By fitting a parabola to the HC and CES sets, we find a degree of curvature of -0.13$_{-0.05}^{+0.05}$\,m\,s$^{-1}$\,yr$^{-2}$. By fitting a parabola only to the HC set, we find a curvature of -0.12$_{-0.05}^{+0.05}$\,m\,s$^{-1}$\,yr$^{-2}$. Therefore the curvature is significant and strongly suggests a Keplerian origin. 

On such a wide orbit, Epsilon Indi Ab is a cold Jovian exoplanet with the longest orbital period among the exoplanets detected through the radial velocity and transit methods.  With components A, Ab, Ba and Bb, Epsilon Indi provides a benchmark system for the formation of gas giants and brown dwarfs. 

\section{Diagnostics of signals using Bayes factor periodograms} \label{sec:signal}
In this section we focus on the constraints provided by the large-scale HARPS dataset alone and investigate evidence for other signals within the the data sets of HH, HL, and HC using the {\small Agatha} software \citep{feng17a}, which is essentially a framework of red noise periodograms. By comparison of different noise models in Agatha, we find that a variety of optimal noise models: (1) HH --- fifth order moving average (or MA(5)) in combination with FWHM and NaD1, (2) HL --- MA(2) combined with BIS and NaD1, (3) HC --- MA(2) combined with S-index and NaD1.

\begin{figure}
  \begin{center}
\includegraphics[scale=0.4]{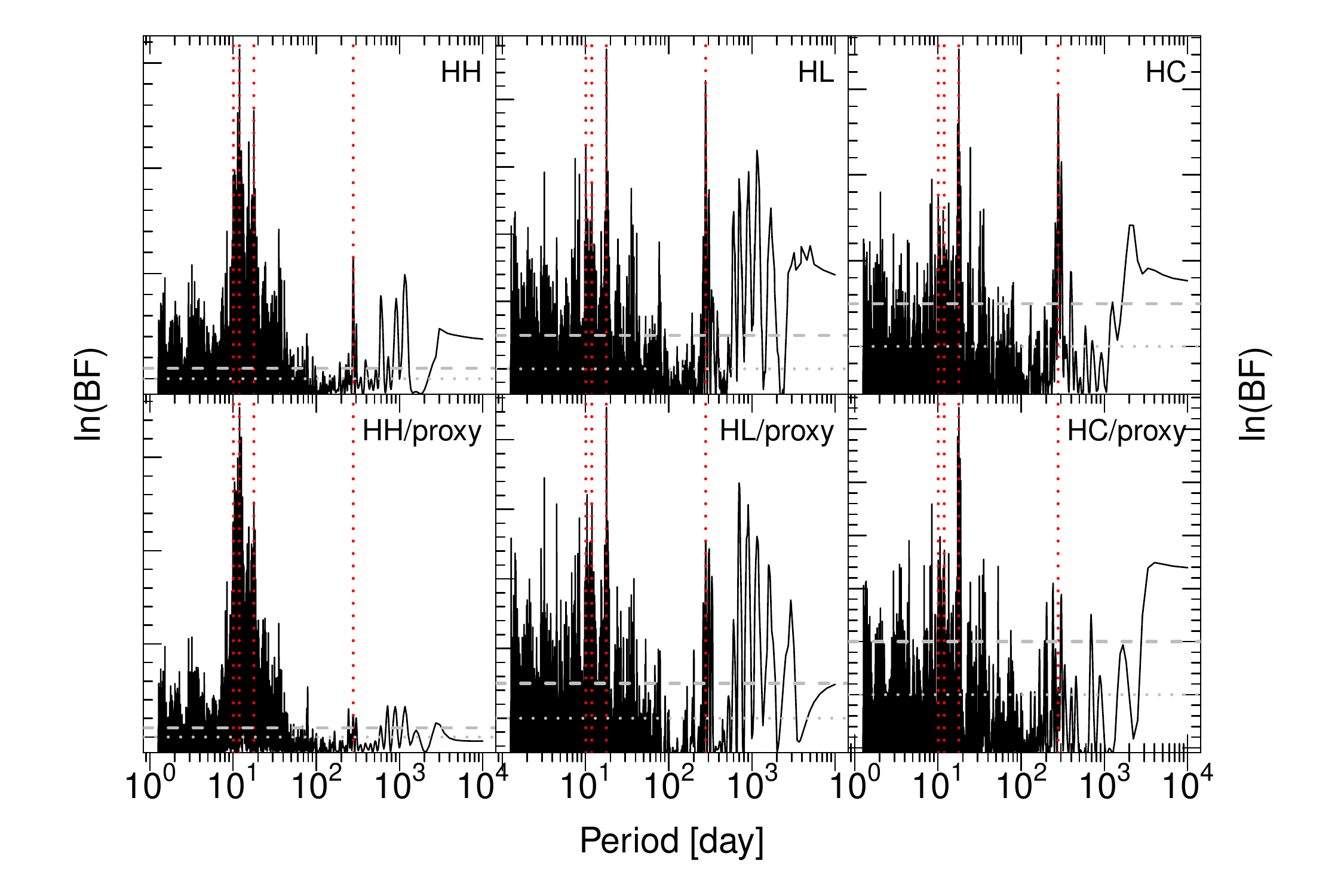}
  \caption{BFPs for MA(1) model in combination with noise proxies for the HH (left), HL (middle) and HC (right) data sets. The upper panels show the BFPs for the MA(1) model while the lowers ones for the MA(1) model in combination with the optimal noise proxies. The signals at periods of 10.1, 11.9, 17.8 and 278\,d are shown by dotted red lines. The thresholds of $\ln({\rm BF})=0$ and 5 are shown by the horizontal dotted and dashed lines, respectively. }
\label{fig:BFP1sig}
\end{center}
\end{figure}

To find primary signals, we calculate the Bayes factor periodogram (BFP) for the MA(1) model in combination with proxies for different data sets, and show them in Fig. \ref{fig:BFP1sig}. In this figure, the signal around 11\,d is significant in the HH data set because the high cadence data favors short period signals. There are also strong powers around this signal for the HL and HC data sets even though the high cadence and postCF epochs have been excluded. For all data sets, the signal at 17.8\,d is significant although the HH data set favors the 11\,d signal more. In particular, the signal at a period of 278\,d is significant in the BFPs for MA(1) but become much weaker in the BFPs for MA(1) combined with noise proxies, suggesting an activity origin. The other strong signals in these BFPs are either aliases or harmonics of these three signals, as we will see in the subsequent analysis. 

\begin{figure*}
  \begin{center}
\includegraphics[scale=0.55]{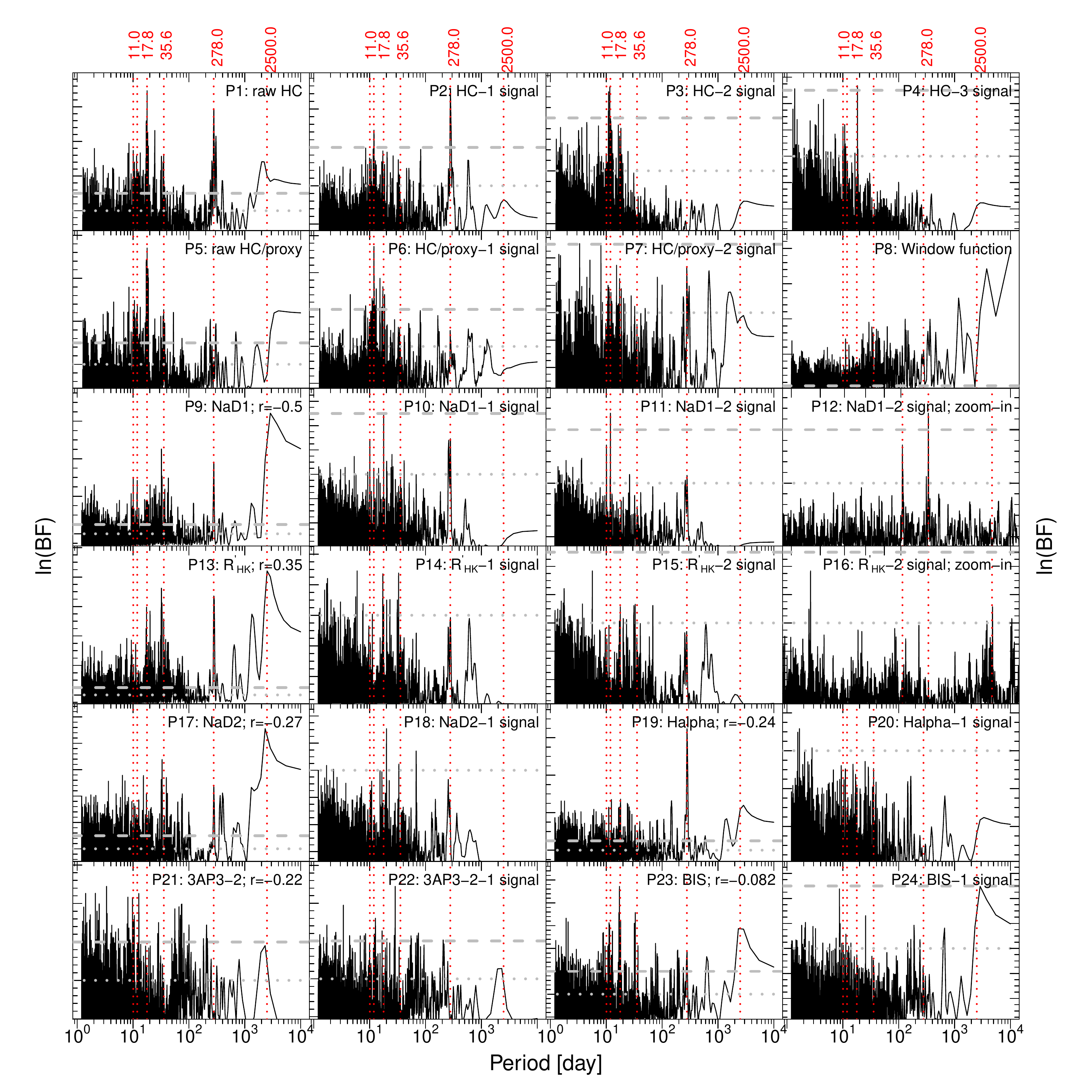}
  \caption{BFPs for RVs and noise proxies. Top row: BFPs for MA(1) for raw HC and HC subsequently subtracted by signals at periods of 11, 17.8, 35.6, 278, and 2500\,d. The 11\,d signal is denoted by two periods at 10.1 and 11.9\,d according to posterior samplings. These periods are shown in the top of the figure. Second row: BFPs for MA(1) combined with S-index and NaD1 for raw HC and HC subtracted in turn by signals at periods at 17.8 and 11\,d, and the Lomb-Scargle periodogram for the window function (P8). The third row onward shows BFPs for MA(1) for noise proxies and their residuals. P12 and P16 are respectively the zoom-in versions of P11 and P15 and have a period range of $[5,~20]$\,d. For each panel, the Pearson correlation coefficient in the top right corner shows the correlation between noise proxies and RVs which are detrended by a second order polynomial. All panels are denoted by ``Pn'' with n varying from 1 to 24. The horizontal dotted and dashed lines denote $\ln{\rm BF}=0$ and 5, respectively. The BFP for the FWHM is not shown because there is not significant periodicity. The values of ln(BF) are indicated by thresholds 0 and 5 rather than shown by axis labels since different panels have different ranges of ln(BF). }
\label{fig:diagnostic}
\end{center}
\end{figure*}

Since the HC data set is more conservative than the others, we calculate the BFPs for the raw HC data and the RV residual signals and for various noise proxies, and show these BFPs in Fig. \ref{fig:diagnostic}. In panels P1-P14, we observed that the signals at periods of 11, 17.8, and 278\,d are significant in the data. The 278\,d signal is likely caused by activity because its significance is reduced after accounting for the correlation between RVs and noise proxies, as seen in P5-P7. This signal is found to be significant in the BFPs of the NaD1, $R_{\rm HK}$, NaD2, and H$\alpha$ indices. In addition, the signal at a period of around 2500\,d is also significant in the BFPs of NaD1, S-index, NaD2, and BIS. We interpret this signal as the magnetic cycle and the 278\,d signal as a secondary cycle in the magnetic variation. We also observe strong signals around 17.8\,d and its double period 35.6\,d in NaD1, NaD2, R$'_{HK}$, and BIS. Nevertheless, the 17.8\,d signal does not disappear in the BFPs accounting for linear correlations between RVs and NaD1 and R$'_{HK}$. Hence we conclude that this signal is either due to a nonlinear effect of stellar rotation or due to a planet, which has an orbital period similar to rotation period. The signal around 11\,d is found to be unique and significant in the residual of NaD1 after subtraction of the 2500 and 17.8\,d signals. Thus this signal is probably caused by stellar activity or is an alias of activity-induced signals. 

Therefore we conclude that the primary and secondary magnetic cycles of Epsilon Indi are 2500 and 278\,d. The rotation period of Epsilon Indi is about 35\,d, approximately double 17.8\,d. This rotation period is rather different from the 22\,d value estimated by \cite{saar97} from Ca II measurements. Considering that the 35\,d rotation period is derived from a relatively large dataset of high precision RVs and multiple activity indicators, we believe that 35\,d is a more reliable value of rotation period. On the other hand, the half rotation period, 17.8\,d, is more significant than the rotation period in the RV data. This phenomenon is also found in the RVs of other stars (e.g. HD 147379; \citealt{feng18d}), and is probably caused by a spot or spot complexity which more significantly modulates spectral lines over one half of the rotation period. The signal at a period of about 11\,d is also related to the stellar rotation since $1/(1/35+1/17.8)=11.8$. Based on the Bayesian quantification of these signals using the MA(1) model \citep{tuomi12,feng16}, the semi-amplitudes of the signals at periods of 10.064$^{+0.003}_{-0.001}$, 17.866$^{+0.006}_{-0.003}$ and 278.0$^{+1.7}_{-0.6}$\,d signals are 1.35$_{-0.31}^{+0.18}$, 2.08$_{-0.19}^{+0.37}$, and 2.28$_{-0.31}^{+0.34}$\,m/s, respectively. The non-detection of additional signals puts an upper limit of 1\,m/s on the semi-amplitude RV variation induced by potential planets in the system. 

\section{Diagnostic procedure in the Agatha framework} \label{sec:diagnose}
In the above section, we find that different noise proxies are sensitive to different stellar activities for Epsilon Indi. NaD1 and R$'_{HK}$ is sensitive to all activity-induced signals while BIS is sensitive to stellar rotation signal. H$\alpha$ is only sensitive to the secondary magnetic cycle. Considering the differential sensitivity of activity indicators to activity signals, a model including a linear correlation between RVs and these indicators would remove activity-noise as well as introduce extra noise, evident from the comparison of P1 with P5 in Fig. \ref{fig:diagnostic}.

\begin{figure*}
 \begin{center}
   \includegraphics[scale=0.45]{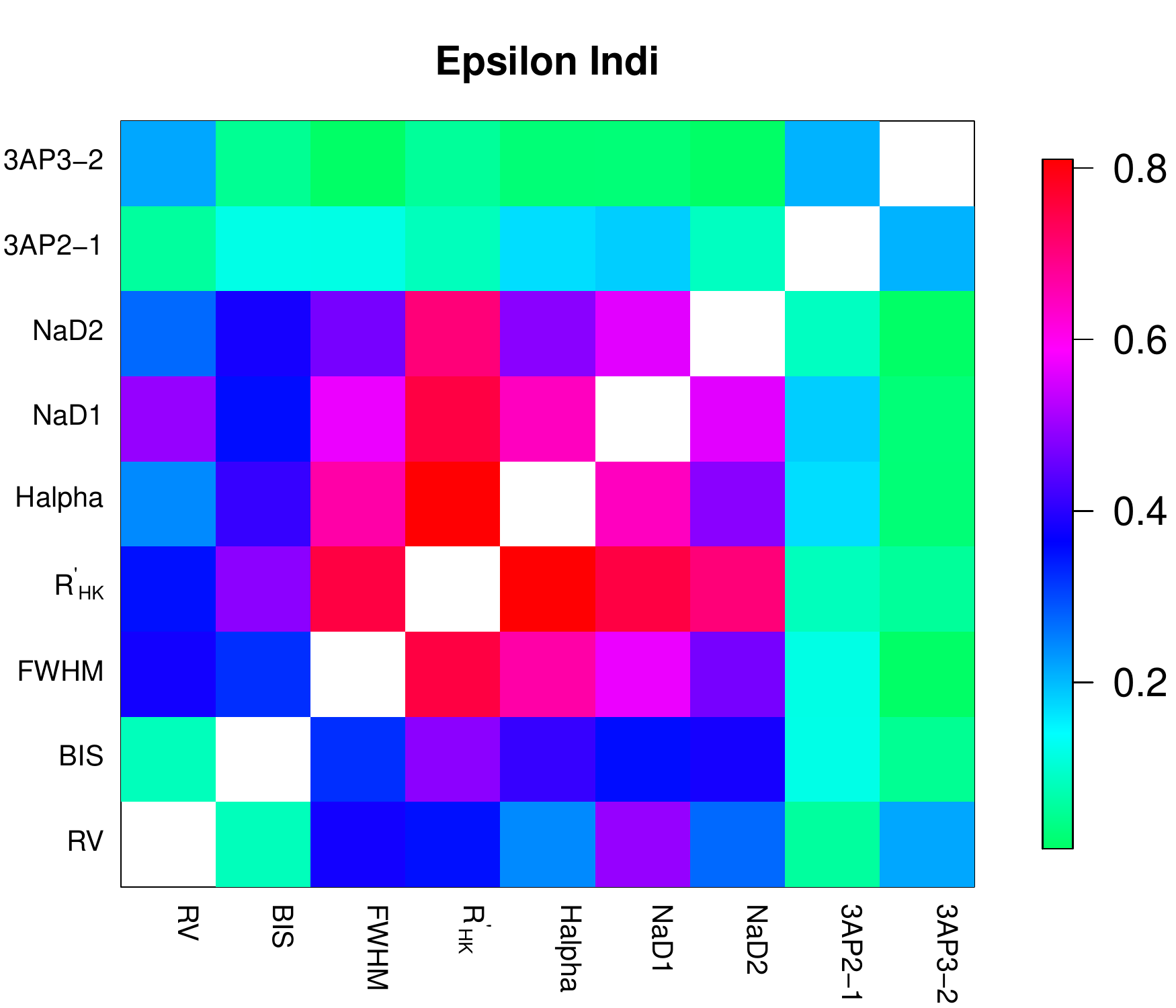}
   \includegraphics[scale=0.45]{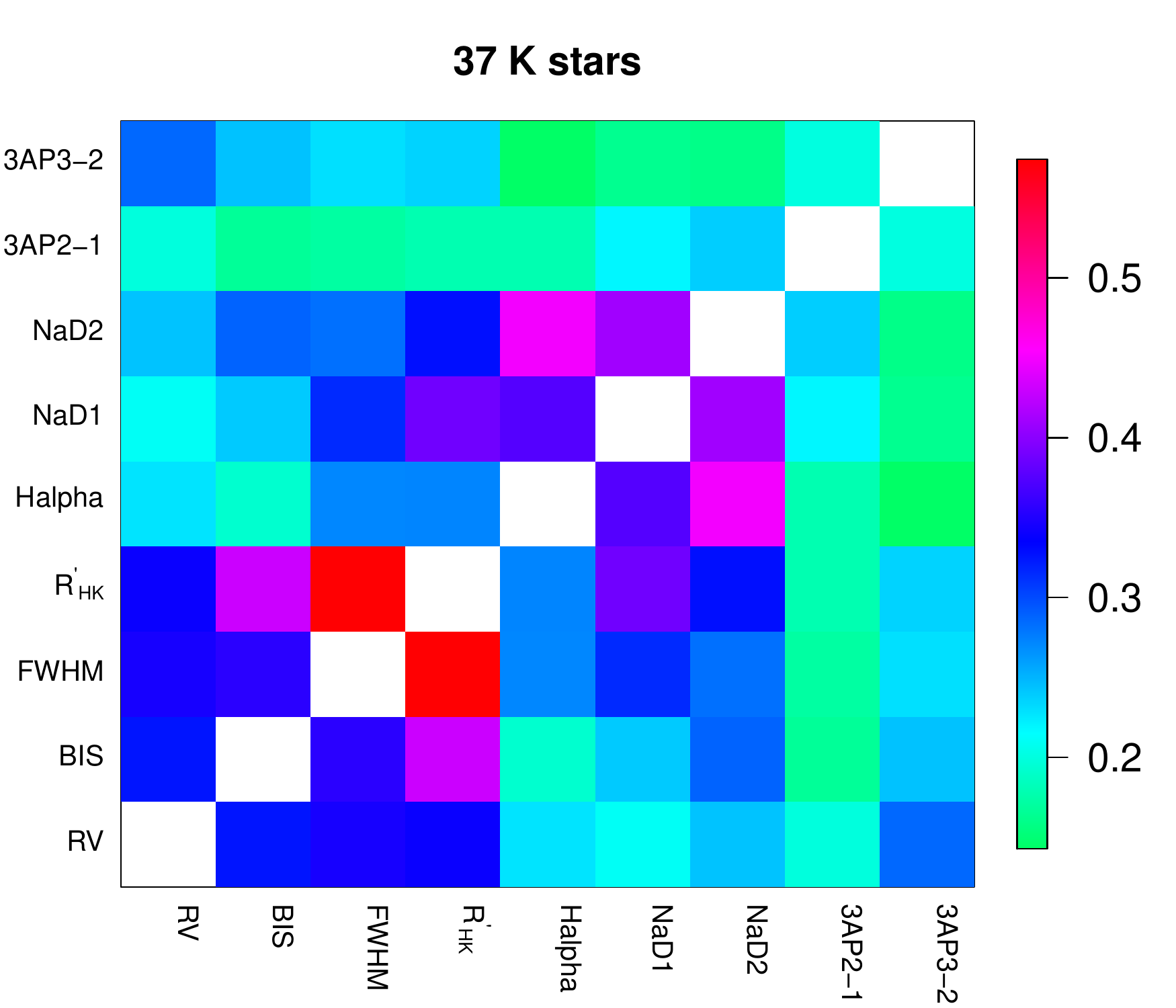}
   \includegraphics[scale=0.45]{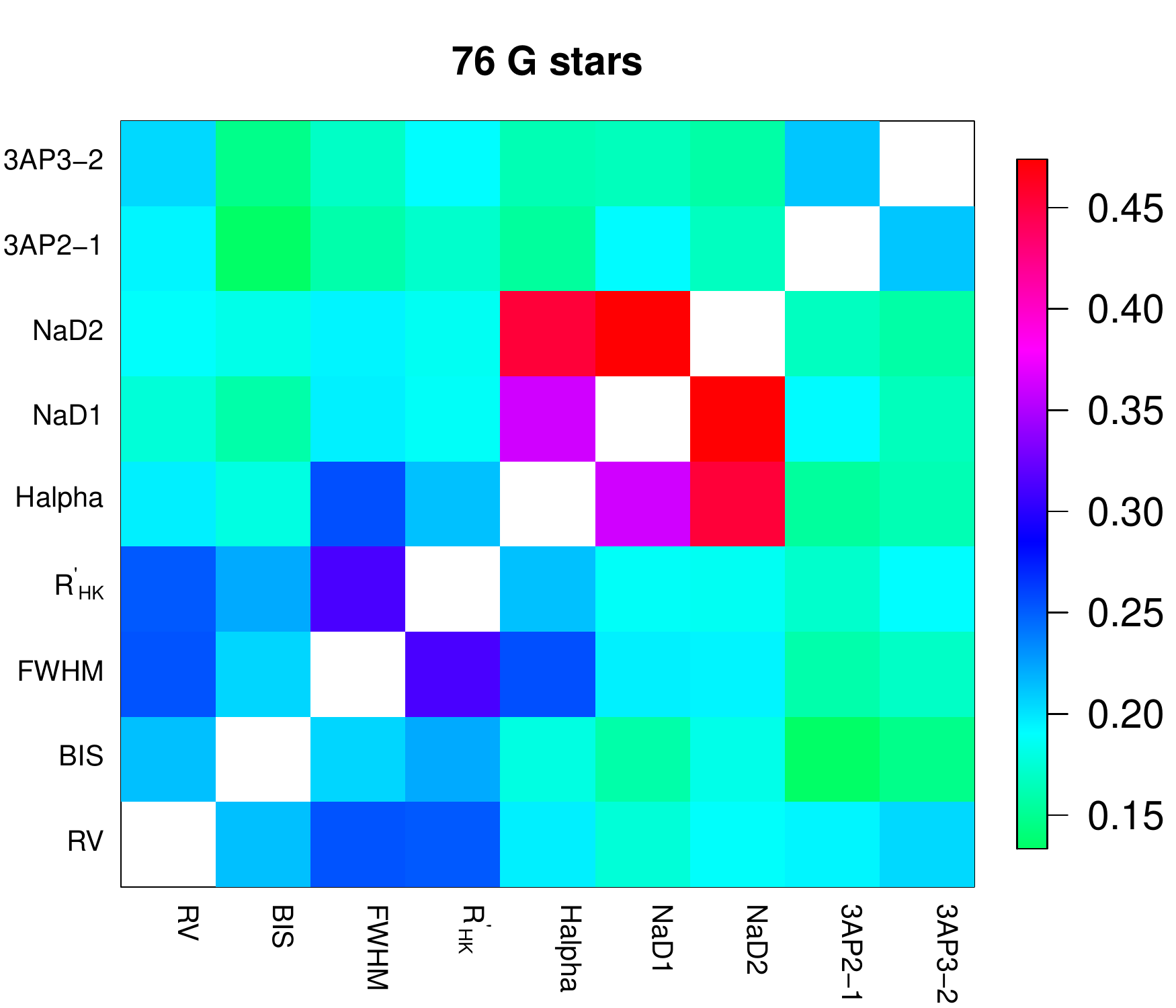}
     \includegraphics[scale=0.45]{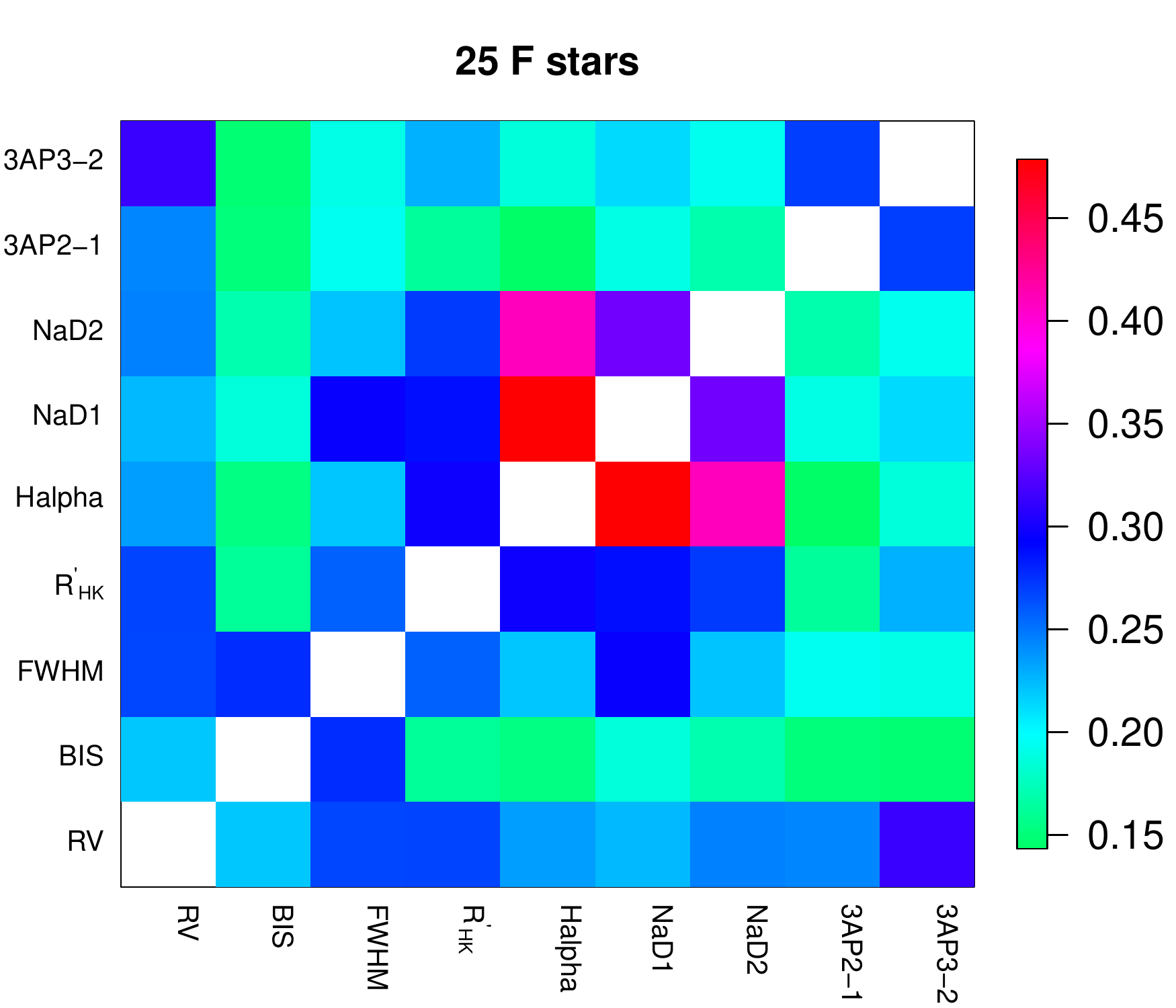}
 \caption{Correlation between detrended RVs and noise proxies for different stellar types. The color encodes the absolute Pearson correlation coefficient. The self-correlation or variance is blank. The number and type of stars is named on top of each panel.}
\label{fig:correlation}
\end{center}
\end{figure*}

Although there is no strong activity-induced signals in the BFPs for differential RV set 3AP3-2, the inclusion of it in the model can efficiently remove wavelength-dependent noise. To investigate this and classify noise proxies, we calculate the correlation between detrended RVs and various noise proxies in the HARPS data sets for Epsilon Indi and for a sample of 138 well observed stars from the ESO HARPS archive\footnote{The HARPS data for these stars are available at the \href{https://github.com/phillippro/agatha/tree/master/data}{Agatha data archive}}, and show them in Fig. \ref{fig:correlation}. For Epsilon Indi, the FWHM,  R$'_{HK}$,  H$\alpha$, NaD1, NaD2,  and 3AP3-2 indices are strongly correlated with RVs. While the other proxies correlate with each other, 3AP3-2 is independent of them. We also see that 3AP3-2 is strongly correlated with RVs for K and F stars but not with G stars, indicating strong wavelength-dependent noise in K and F stars. For all types of stars, 3AP3-2 is not correlated with traditional activity indices especially the H$\alpha$, NaD1, and NaD2 indices. This is reasonable since these indices are determined from single spectral lines and thus are not wavelength-dependent. We also observe that the sodium lines (NaD1 and NaD2) strongly correlate with H$\alpha$ for all types of stars, especially for G type stars. On the other hand, R$'_{HK}$ is not strongly correlated with H$\alpha$ for any type of stars. This apparently contradicts the conclusion of \cite{silva13} who found a strong correlation between R$'_{HK}$ and $\log~{\rm H}\alpha$ albeit for a different sample of stars and of course we find a strong correlation for Epsilon Indi A. Although NaD1 strongly correlates with RVs and other indices for Epsilon Indi, we do not see a strong averaged correlation in the sample of stars. The strong correlation between RVs and NaD1 is also found for M dwarfs by \citep{robertson15}. But we do not investigate this in detail due to a relative lack of HARPS data for M dwarfs. We also find strong correlation between FWHM, R$'_{HK}$, and RVs for all types of stars, especially K stars. The BIS only strongly correlates with RVs for K type stars. 

Based on these observations, we conclude that differential RVs are an important complement to traditional activity indicators in terms of diagnosing and modeling wavelength-dependent activity signals in RV data. To diagnose activity signals properly, we suggest the following {\small Agatha}-based activity diagnostic sequence for the analysis of a given RV data set.
\begin{itemize}
\item Compare noise models and select the optimal moving average model and noise proxies \citep{feng17a}
\item Calculate the BFPs for the optimal red noise model (MA(1) is recommended to reduce computation time) for RVs and RV residuals by subtracting signals 
\item Repeat the first step but including the optimal noise proxies into the red noise model
\item Repeat the first step but including the differential RVs into the red noise model
\item If all signals passing the Bayes factor threshold of $\ln({\rm BF})>5$ \citep{feng16} are sensitive to the choice of noise models, stop the diagnostics and conclude that all signals are induced by activity and there is no detectable planetary signal in the data
\item If some signals are insensitive to the choice of noise models, calculate BFPs for noise proxies and their residuals to find potential overlaps between RV signals and proxy signals. 
\item Check the consistency of signals with time using the moving periodogram introduced by \cite{feng17a}.
\item If any signals are insensitive to the choice of noise models, and are not found in noise proxies, and are consistent with time, interpret them as planetary candidates and quantify them using MCMC samplings. Otherwise, conclude that all signals are caused by activity. 
\end{itemize}
The moving periodogram is not used in this work because the BFP-based diagnostics already exclude the Keplerian origin of any signals.

\section{Discussions and conclusion}\label{sec:conclusion}
We analyze the HARPS data of Epsilon Indi A in the {\small Agatha} framework in combination with Bayesian methods. We confirm the suspected planet Epsilon Indi Ab to be a cold Jovian planet on a 11.91\,au-wide orbit with an orbital period of 52.62\,yr, making it the planet with the longest period from exoplanets detected by the RV methods. Epsilon Indi Ab is only 3.62\,pc away from the Sun and is the closest Jovian exoplanet from the Earth. Given its proximity to the Sun, Epsilon Indi Ab is separated from its host star by 3.3\,'' and thus observed through direction imaging and astrometry, for example by JWST and Gaia \citep{krist07,perryman14}.

With one Jovian planet and a binary brown dwarf system, it provides a benchmark case to test theories on the formation of giant planets and brown dwarfs. According to \cite{kennedy08}, stars less massive than the Sun are unlikely to form more than one giant planet. Thus Ba and Bb are more likely to be captured by Epsilon Indi A either in or out of its birth environment. To investigate this, we calculate the escape radius derived by \cite{feng17c} based on simulations of perturbations from stellar encounters on wide binaries. Adopting an age of 1.4\,Gyr \citep{bonfanti14} and an encounter rate of 80\,Myr$^{-1}$ for stellar encounters with periapsis less than 1\,pc, the escape radius of Epsilon Indi A is about 11000\,au which is well beyond the projected separation (1459\,au) between Epsilon Indi B and A \citep{scholz03}. However, if Epsilon Indi A migrated outward to its current location or captured Epsilon Indi B during its formation, the encounter rate would be much higher and the escaped radius could be within the orbit of Epsilon Indi B, which may also be much larger than the projected separation due to geometric effect. Thus there is considerable evidence for the capture scenario for the formation of Epsilon Indi, like the Proxima-alpha Centauri system \citep{feng18a}. In addition, the lack of planets with $K>1$\,m/s suggests a lack of super-Earths/Neptunes in the habitable zone (0.47 to 0.86\,au; \citealt{kopparapu14}) of Epsilon Indi A. Thus only Earth-like or smaller planets are allowed in the habitable zone, consistent with the conclusion in \cite{seppeur18} that K dwarfs are more likely to host dynamically stable habitable planets than more massive stars if the outside giant planet is not on an eccentric orbit. If these rocky planets are detected, Epsilon Indi would be similar to the Solar System, with close-in rocky planets and longer period gas giants. 

We also diagnose the other signals by comparing the BFPs for various data sets, noise models, and noise proxies. We find three signals at periods of about 11, 17.8 and 278\,d in the RV data of Epsilon Indi A. These signals are significant and can be constrained by Bayesian posterior samplings. Nevertheless, they are unlikely to be Keplerian because significant powers around these signals are found in the periodograms of various noise proxies especially in sodium lines. Based on the activity diagnostics, we conclude that all these signals together with the 2500\,d signal in many noise proxies are caused by stellar activity. In particular, the 2500 and 278\,d signals correspond to the magnetic cycles of Epsilon Indi A while the 35, 17.8 and 11\,d signals are related to the stellar rotation. 

We find that the correlation between RVs and activity indicators depend on activity time scales. Thus the inclusion of them in the fit would remove activity-induced noise as well as introduce extra noise. While the traditional activity indicators are limited in removing activity-induced signals, differential RVs are better in removing wavelength-dependent signals without introducing extra noise because they provide a simple way to weight spectral orders {\it a posteriori}. Thus differential RVs are an important complement to the traditional activity indicators in activity diagnostics. 

Since the signals in the raw data are quite significant, it would be easy to misinterpret them as planet candidates without careful activity diagnostics. To reliably detect low amplitude signals, we propose a diagnostic procedure in the {\small Agatha} framework. This procedure is able to identify activity-induced signals through the combined diagnostics of the dependence of signals on wavelength, activity indices, and time. The lesson we learn from the analysis of the HARPS data for Epsilon Indi A would benefit the detection of Earth analogs in high precision RV data, for example, from ESPRESSO \citep{hernandez17}. 

\section*{Acknowledgements}
This work is supported by the Science and Technology Facilities Council [ST/M001008/1]. We used the ESO Science Archive Facility to collect radial velocity data. 
\bibliographystyle{aasjournal}
\bibliography{nm}

\begin{thebibliography}{14}
\expandafter\ifx\csname natexlab\endcsname\relax\def\natexlab#1{#1}\fi

\bibitem[{{Anglada-Escud{\'e}} {et~al}\mbox{.}(2016){Anglada-Escud{\'e}},
  {Amado}, {Barnes}, {Berdi{\~n}as}, {Butler}, {Coleman}, {de La Cueva},
  {Dreizler}, {Endl}, {Giesers}, {Jeffers}, {Jenkins}, {Jones}, {Kiraga},
  {K{\"u}rster}, {L{\'o}pez-Gonz{\'a}lez}, {Marvin}, {Morales}, {Morin},
  {Nelson}, {Ortiz}, {Ofir}, {Paardekooper}, {Reiners}, {Rodr{\'{\i}}guez},
  {Rodr{\'{\i}}guez-L{\'o}pez}, {Sarmiento}, {Strachan}, {Tsapras}, {Tuomi}, \&
  {Zechmeister}}]{anglada16}
{Anglada-Escud{\'e}} G. {et~al.}, 2016, \nat, 536, 437

\bibitem[{{Anglada-Escud{\'e}} \& {Butler}(2012)}]{anglada12}
{Anglada-Escud{\'e}} G., {Butler} R.~P., 2012, \apjs, 200, 15

\bibitem[{{Bonfanti} {et~al}\mbox{.}(2015){Bonfanti}, {Ortolani}, {Piotto}, \&
  {Nascimbeni}}]{bonfanti14}
{Bonfanti} A. {et~al.}, 2015, \aap, 575, A18

\bibitem[{{Demory} {et~al}\mbox{.}(2009){Demory}, {S{\'e}gransan}, {Forveille},
  {Queloz}, {Beuzit}, {Delfosse}, {di Folco}, {Kervella}, {Le Bouquin},
  {Perrier}, {Benisty}, {Duvert}, {Hofmann}, {Lopez}, \& {Petrov}}]{demory09}
{Demory} B.-O. {et~al.}, 2009, \aap, 505, 205

\bibitem[{{Endl} {et~al}\mbox{.}(2002){Endl}, {K{\"u}rster}, {Els}, {Hatzes},
  {Cochran}, {Dennerl}, \& {D{\"o}bereiner}}]{endl02}
{Endl} M. {et~al.}, 2002, \aap, 392, 671

\bibitem[{{Feng}, {Tuomi} \& {Jones}(2017{\natexlab{a}}){Feng}, {Tuomi}, \&
  {Jones}}]{feng17a}
{Feng} F., {Tuomi} M., {Jones} H.~R.~A., 2017{\natexlab{a}}, \mnras, 470, 4794

\bibitem[{{Feng}, {Tuomi} \& {Jones}(2017{\natexlab{b}}){Feng}, {Tuomi}, \&
  {Jones}}]{feng17b}
{Feng} F., {Tuomi} M., {Jones} H.~R.~A., 2017{\natexlab{b}}, \aap, 605, A103

\bibitem[{{Feng} {et~al}\mbox{.}(2016){Feng}, {Tuomi}, {Jones}, {Butler}, \&
  {Vogt}}]{feng16}
{Feng} F. {et~al.}, 2016, \mnras, 461, 2440

\bibitem[{{Gonz{\'a}lez Hern{\'a}ndez} {et~al}\mbox{.}(2017){Gonz{\'a}lez
  Hern{\'a}ndez}, {Pepe}, {Molaro}, \& {Santos}}]{hernandez17}
{Gonz{\'a}lez Hern{\'a}ndez} J.~I. {et~al.}, 2017, ArXiv e-prints

\bibitem[{{Lingam} \& {Loeb}(2017)}]{lingam17}
{Lingam} M., {Loeb} A., 2017, ArXiv e-prints

\bibitem[{{Scholz} {et~al}\mbox{.}(2003){Scholz}, {McCaughrean}, {Lodieu}, \&
  {Kuhlbrodt}}]{scholz03}
{Scholz} R.-D. {et~al.}, 2003, \aap, 398, L29

\bibitem[{{Tuomi} {et~al}\mbox{.}(2013){Tuomi}, {Jones}, {Jenkins}, {Tinney},
  {Butler}, {Vogt}, {Barnes}, {Wittenmyer}, {O'Toole}, {Horner}, {Bailey},
  {Carter}, {Wright}, {Salter}, \& {Pinfield}}]{tuomi12}
{Tuomi} M. {et~al.}, 2013, \aap, 551, A79

\bibitem[{{van Leeuwen}(2007)}]{leeuwen07}
{van Leeuwen} F., 2007, \aap, 474, 653

\bibitem[{{Zechmeister} {et~al}\mbox{.}(2013){Zechmeister}, {K{\"u}rster},
  {Endl}, {Lo Curto}, {Hartman}, {Nilsson}, {Henning}, {Hatzes}, \&
  {Cochran}}]{zechmeister13}
{Zechmeister} M. {et~al.}, 2013, \aap, 552, A78

\end{thebibliography}


\begin{thebibliography}{}
\expandafter\ifx\csname natexlab\endcsname\relax\def\natexlab#1{#1}\fi

\bibitem[{{Anglada-Escud{\'e}} \& {Butler}(2012)}]{anglada12}
{Anglada-Escud{\'e}}, G., \& {Butler}, R.~P. 2012, \apjs, 200, 15

\bibitem[{{Anglada-Escud{\'e}} {et~al.}(2016){Anglada-Escud{\'e}}, {Amado},
  {Barnes}, {Berdi{\~n}as}, {Butler}, {Coleman}, {de La Cueva}, {Dreizler},
  {Endl}, {Giesers}, {Jeffers}, {Jenkins}, {Jones}, {Kiraga}, {K{\"u}rster},
  {L{\'o}pez-Gonz{\'a}lez}, {Marvin}, {Morales}, {Morin}, {Nelson}, {Ortiz},
  {Ofir}, {Paardekooper}, {Reiners}, {Rodr{\'{\i}}guez},
  {Rodr{\'{\i}}guez-L{\'o}pez}, {Sarmiento}, {Strachan}, {Tsapras}, {Tuomi}, \&
  {Zechmeister}}]{anglada16}
{Anglada-Escud{\'e}}, G., {Amado}, P.~J., {Barnes}, J., {et~al.} 2016, \nat,
  536, 437

\bibitem[{{Bonfanti} {et~al.}(2015){Bonfanti}, {Ortolani}, {Piotto}, \&
  {Nascimbeni}}]{bonfanti14}
{Bonfanti}, A., {Ortolani}, S., {Piotto}, G., \& {Nascimbeni}, V. 2015, \aap,
  575, A18

\bibitem[{{Demory} {et~al.}(2009){Demory}, {S{\'e}gransan}, {Forveille},
  {Queloz}, {Beuzit}, {Delfosse}, {di Folco}, {Kervella}, {Le Bouquin},
  {Perrier}, {Benisty}, {Duvert}, {Hofmann}, {Lopez}, \& {Petrov}}]{demory09}
{Demory}, B.-O., {S{\'e}gransan}, D., {Forveille}, T., {et~al.} 2009, \aap,
  505, 205

\bibitem[{{Endl} {et~al.}(2002){Endl}, {K{\"u}rster}, {Els}, {Hatzes},
  {Cochran}, {Dennerl}, \& {D{\"o}bereiner}}]{endl02}
{Endl}, M., {K{\"u}rster}, M., {Els}, S., {et~al.} 2002, \aap, 392, 671

\bibitem[{{Feng} \& {Jones}(2018)}]{feng18a}
{Feng}, F., \& {Jones}, H.~R.~A. 2018, \mnras, 473, 3185

\bibitem[{{Feng} {et~al.}(2018){Feng}, {Jones}, \& {Tuomi}}]{feng18d}
{Feng}, F., {Jones}, H.~R.~A., \& {Tuomi}, M. 2018, Research Notes of the
  American Astronomical Society, 2, 23

\bibitem[{{Feng} {et~al.}(2017{\natexlab{a}}){Feng}, {Tuomi}, \&
  {Jones}}]{feng17a}
{Feng}, F., {Tuomi}, M., \& {Jones}, H.~R.~A. 2017{\natexlab{a}}, \mnras, 470,
  4794

\bibitem[{{Feng} {et~al.}(2017{\natexlab{b}}){Feng}, {Tuomi}, \&
  {Jones}}]{feng17b}
---. 2017{\natexlab{b}}, \aap, 605, A103

\bibitem[{{Feng} {et~al.}(2017{\natexlab{c}}){Feng}, {Tuomi}, {Jones},
  {Barnes}, {Anglada-Escud{\'e}}, {Vogt}, \& {Butler}}]{feng17c}
{Feng}, F., {Tuomi}, M., {Jones}, H.~R.~A., {et~al.} 2017{\natexlab{c}}, \aj,
  154, 135

\bibitem[{{Feng} {et~al.}(2016){Feng}, {Tuomi}, {Jones}, {Butler}, \&
  {Vogt}}]{feng16}
{Feng}, F., {Tuomi}, M., {Jones}, H.~R.~A., {Butler}, R.~P., \& {Vogt}, S.
  2016, \mnras, 461, 2440

\bibitem[{{Gomes da Silva} {et~al.}(2014){Gomes da Silva}, {Santos}, {Boisse},
  {Dumusque}, \& {Lovis}}]{silva13}
{Gomes da Silva}, J., {Santos}, N.~C., {Boisse}, I., {Dumusque}, X., \&
  {Lovis}, C. 2014, \aap, 566, A66

\bibitem[{{Gonz{\'a}lez Hern{\'a}ndez} {et~al.}(2017){Gonz{\'a}lez
  Hern{\'a}ndez}, {Pepe}, {Molaro}, \& {Santos}}]{hernandez17}
{Gonz{\'a}lez Hern{\'a}ndez}, J.~I., {Pepe}, F., {Molaro}, P., \& {Santos}, N.
  2017, ArXiv e-prints, arXiv:1711.05250

\bibitem[{Haario {et~al.}(2006)Haario, Laine, Mira, \& Saksman}]{haario06}
Haario, H., Laine, M., Mira, A., \& Saksman, E. 2006, Statistics and Computing,
  16, 339

\bibitem[{{Janson} {et~al.}(2009){Janson}, {Apai}, {Zechmeister}, {Brandner},
  {K{\"u}rster}, {Kasper}, {Reffert}, {Endl}, {Lafreni{\`e}re}, {Gei{\ss}ler},
  {Hippler}, \& {Henning}}]{janson09}
{Janson}, M., {Apai}, D., {Zechmeister}, M., {et~al.} 2009, \mnras, 399, 377

\bibitem[{{Kennedy} \& {Kenyon}(2008)}]{kennedy08}
{Kennedy}, G.~M., \& {Kenyon}, S.~J. 2008, \apj, 673, 502

\bibitem[{{Kopparapu} {et~al.}(2014){Kopparapu}, {Ramirez}, {SchottelKotte},
  {Kasting}, {Domagal-Goldman}, \& {Eymet}}]{kopparapu14}
{Kopparapu}, R.~K., {Ramirez}, R.~M., {SchottelKotte}, J., {et~al.} 2014,
  \apjl, 787, L29

\bibitem[{{Krist} {et~al.}(2007){Krist}, {Beichman}, {Trauger}, {Rieke},
  {Somerstein}, {Green}, {Horner}, {Stansberry}, {Shi}, {Meyer}, {Stapelfeldt},
  \& {Roellig}}]{krist07}
{Krist}, J.~E., {Beichman}, C.~A., {Trauger}, J.~T., {et~al.} 2007, in
  \procspie, Vol. 6693, Techniques and Instrumentation for Detection of
  Exoplanets III, 66930H

\bibitem[{{Lingam} \& {Loeb}(2017)}]{lingam17}
{Lingam}, M., \& {Loeb}, A. 2017, ArXiv e-prints, arXiv:1703.00878

\bibitem[{{Lo Curto} {et~al.}(2015){Lo Curto}, {Pepe}, {Avila}, {Boffin},
  {Bovay}, {Chazelas}, {Coffinet}, {Fleury}, {Hughes}, {Lovis}, {Maire},
  {Manescau}, {Pasquini}, {Rihs}, {Sinclaire}, \& {Udry}}]{curto15}
{Lo Curto}, G., {Pepe}, F., {Avila}, G., {et~al.} 2015, The Messenger, 162, 9

\bibitem[{{Perryman} {et~al.}(2014){Perryman}, {Hartman}, {Bakos}, \&
  {Lindegren}}]{perryman14}
{Perryman}, M., {Hartman}, J., {Bakos}, G.~{\'A}., \& {Lindegren}, L. 2014,
  \apj, 797, 14

\bibitem[{{Robertson} {et~al.}(2015){Robertson}, {Endl}, {Henry}, {Cochran},
  {MacQueen}, \& {Williamson}}]{robertson15}
{Robertson}, P., {Endl}, M., {Henry}, G.~W., {et~al.} 2015, \apj, 801, 79

\bibitem[{{Saar} \& {Osten}(1997)}]{saar97}
{Saar}, S.~H., \& {Osten}, R.~A. 1997, \mnras, 284, 803

\bibitem[{{Scholz} {et~al.}(2003){Scholz}, {McCaughrean}, {Lodieu}, \&
  {Kuhlbrodt}}]{scholz03}
{Scholz}, R.-D., {McCaughrean}, M.~J., {Lodieu}, N., \& {Kuhlbrodt}, B. 2003,
  \aap, 398, L29

\bibitem[{{Seppeur}(2018)}]{seppeur18}
{Seppeur}, S. 2018, ArXiv e-prints, arXiv:1802.05736

\bibitem[{{Tuomi} {et~al.}(2013){Tuomi}, {Jones}, {Jenkins}, {Tinney},
  {Butler}, {Vogt}, {Barnes}, {Wittenmyer}, {O'Toole}, {Horner}, {Bailey},
  {Carter}, {Wright}, {Salter}, \& {Pinfield}}]{tuomi12}
{Tuomi}, M., {Jones}, H.~R.~A., {Jenkins}, J.~S., {et~al.} 2013, \aap, 551, A79

\bibitem[{{van Leeuwen}(2007)}]{leeuwen07}
{van Leeuwen}, F. 2007, \aap, 474, 653

\bibitem[{{Zechmeister} {et~al.}(2013){Zechmeister}, {K{\"u}rster}, {Endl}, {Lo
  Curto}, {Hartman}, {Nilsson}, {Henning}, {Hatzes}, \&
  {Cochran}}]{zechmeister13}
{Zechmeister}, M., {K{\"u}rster}, M., {Endl}, M., {et~al.} 2013, \aap, 552, A78

\end{thebibliography}
\end{document}